\documentstyle[aps,prl]{revtex}

\begin{document}
\title{The de Haas - van Alphen effect in two-dimensional metals.}
\author{P.D.Grigoriev$^{1,2}$, I.D.Vagner$^{1,3}$}
\date{\today}
\address{$^{1}$Grenoble High Magnetic Field Laboratory \\
MPI-FKF and CNRS \\
BP 166, F-38042 Grenoble Cedex 09, France\\
$^{2}$L.D.Landau Institute for Theoretical Physics \\
142432, Chernogolovka, Moscow region, Russia \\
e-mail: pashag@itp.ac.ru\\
$^{3}$Physics and Engineering Institute at Ruppin ,\\
Emek Hefer 40250, Israel\\
}
\maketitle

\begin{abstract}
The de Haas - van Alphen effect in two-dimensional (2D) metals is
investigated at different conditions and with different shapes of Landau
levels (LLs). The analytical calculations can be done when many LLs are
occupied. We consider the cases of fixed particle number ($N=const$), fixed
chemical potential ($\mu =const$) and the intermediate situation of finite
electron reservoir. The last case takes place in organic metals due to
quasi-one-dimensional sheets of Fermi surface.

We obtained the envelopes of magnetization oscillations in all these cases
in the limit of low temperature and Dingle temperature, where the
oscillations can not be approximated by few first terms in the harmonic
expansion. The results are compared and shown to be substantially different
for different shapes of LLs.

The simple relation between the shape of LLs and the wave form of
magnetization oscillations is found. It allows to obtain the density of
states distribution at arbitrary magnetic field and spin-splitting using the
measurement of the magnetization curve.

The analytical formula for the magnetization at $\mu =const$ and the
Lorentzian shape of LLs at arbitrary temperature, Dingle temperature and
spin splitting is obtained and used to examine the possibility of the
diamagnetic phase transition in 2D metals.
\end{abstract}

\bigskip PACS: \bigskip

\section{Introduction}

The quantum oscillations of magnetization and conductivity in magnetic field
(the de Haas - van Alphen and Shubnikov - de Haas effects) give abundant
information about electronic structure of metals \cite{S0}. The theory of
these effects in three-dimensional (3D) metals has been constructed \cite{LK}
and proved well by many experiments \cite{S0}. However, this theory is not
applicable in two-dimensional (2D) compounds, when warping of the Fermi
surface(FS) is less than the Landau level separation $\hbar \omega _{c}$.
The calculation of the kinetic properties of 2D compounds in magnetic field
is difficult even in the simplest model approximations \cite{TA},\cite{M}.
Including of the electron-electron interaction, other parts of FS or
different scattering mechanisms makes the problem formidable and the
quantitative theory of the 2D Shubnikov - de Haas effect does not exist. The
theory of the de Haas - van Alphen (dHvA) effect is much simpler because the
magnetization is a thermodynamic quantity. The complete description of the
2D dHvA effect, however, does not exist either. There are two approaches to
this problem. The first is to make the harmonic expansion of magnetization
oscillations\cite{Sh}, in analogy to the Lifshitz-Kosevich formula \cite{LK}%
. This expansion assumes the chemical potential to be fixed ($\mu =const$).
In most compounds it is the electron density which is constant and the
chemical potential is an oscillating function of the magnetic field. Even at
fixed chemical potential this approach in pure metals at low temperature
gives the infinite harmonic series for magnetization oscillations with slow
convergence which is not convenient for the practical use. This series has
been summed explicitly only at zero temperature and the Lorentzian shape of
LLs \cite{Min}.

Another approach is the direct summation over Landau levels (LLs). It is
applicable to the systems with fixed electron density ($N=const$) and just
at low temperature $T\ll \hbar \omega _{c}$ ($\hbar \omega _{c}=\hbar
eB/m^{\ast }c$ is the LL separation) since then only few LLs near FS are
important\cite{V1}. The analytical calculations of the magnetization and its
envelope by this method at low but nonzero temperature have been done \cite
{V1},\cite{GV} only for sharp Landau levels(LLs). This approximation has a
very narrow domain of applicability $\hbar \omega _{c}\gg T\gg \sqrt{\Gamma
\hbar \omega _{c}}$ because the Landau level broadening $\Gamma $ has a very
strong influence on the magnetization oscillations. Any electron reservoirs
are also ignored in these calculations although in organic metal this
reservoir of field independent electron states exists due to
quasi-one-dimensional parts of FS. The influence of this electron reservoir
on the magnetization oscillations has been studied recently \cite{IMV} only
for sharp LLs. Numerical calculations are possible \cite{H},\cite{F}, but
the numerical analysis is not evident and convenient for practical use. So,
in many cases one continues to apply the Shoenberg's 2D harmonic expansion.

Our work is intended to study the 2D dHvA effect with arbitrary shape and
width of the LLs and under conditions close to real experiments.

In sections II, III, IV we consider the case of fixed electron density using
the method of direct summation over LLs. The reservoir electron states and
finite temperature are also taken into account. In Sec. II we derive the
equation for the chemical potential and the relation between the
magnetization and the chemical potential that are valid at arbitrary shape
and width of LLs. In section III this is used to derive the simple relation
between magnetization oscillations and the density of states (DoS) function.
We describe in details how this relationship enables the direct measurement
of the DoS distribution.

The problem of the density of states distribution in 2D metals in magnetic
field is an important and old one. Since the discovery of the integer
quantum Hall effect many theoretical approaches have been developed to solve
it (for example, \cite{TA},\cite{M}) and it would be very interesting to
compare the results of these calculations with the experimentally measured
DoS distribution. The DoS function tells about the electron spectrum and the
scattering mechanisms that determine the kinetic phenomena but it is
difficult to explicitly relate the DoS distribution to the conductivity. The
magnetization gives a quite direct information about the DoS and an accurate
method to extract the DoS distribution from the magnetization oscillations
is needed. The harmonic expansion is not good here for the following
reasons. To obtain the DoS distribution function one should consider the
case of constant electron density and low temperature when many harmonics
are important. In this case the explicit harmonic expansion (with many
terms) of magnetization oscillations does not exist and the influence of LL
broadening on the harmonic damping can not be separated accurately from
other factors. The method of extracting the DoS distribution from the dHvA
measurements has been proposed by M. Ya. Azbel\cite{Az}. This method is
valid only in the extremely strong magnetic field when the magnetic length $%
l_{0}=(c\hbar /eB)^{1/2}$ \ is much less than the impurity Bohr radius. His
procedure assumes that one can realize the situation when only one LL is
occupied (this means that $\hbar \omega _{c}/2>E_{F}$ , $E_{F}$ is the Fermi
energy). Such a condition can not be fulfilled in organic metals and many
other compounds where Fermi energy is quite great. Azbel's procedure also
assumes zero temperature, no electron reservoir and no warping of Fermi
surface. Our approach has no these defaults and limitations.

In Sec. IV we obtain the analytical formulas for the envelopes of the
magnetization oscillations. In Sec. V we study the case of constant chemical
potential $\mu =const$ using the harmonic expansion. We obtain the
analytical formula for the magnetization with Lorentzian shape of LLs for
arbitrary temperature, Dingle temperature and spin splitting. This formula
is used to examine the possibility of the diamagnetic phase transition in 2D
metals. We also derive the relations between the oscillations of the
magnetization and the oscillations of the chemical potential (in the case $%
N=const$) or the particle density (when $\mu =const$) at general conditions
and find their domain of applicability.

Both the conditions $N=const$ and $\mu =const$ take place in real compounds
that was determined on the experiment using the shape of magnetization
oscillations \cite{Wieg},\cite{IV},\cite{Fix} or by the observation of $%
\beta -\alpha $ frequency \cite{W},\cite{Z} in the presence of the magnetic
breakdown. Therefore we consider both these cases.

\section{Derivation of magnetization in the case N=const with finite
electron reservoir}

The condition of the constant particle density ($N=const$) means that the
sample is isolated from electron reservoirs and the sample volume is
constant. This takes place in heterostructures \cite{Wieg} and is usually
more realistic than restriction of the fixed chemical potential ($\mu =const$%
). Nevertheless, in many solids the exact conditions $N=const$ or $\mu
=const $ are not fulfilled. The intermediate situation takes place, for
example, in synthetic metals like intercalated graphite compounds where the
magnetic field dependent charge transfer between the graphite layers and
intercalated layers takes place, and in organic metals due to the additional
quasi-1D parts of FS, that does not contribute to the magnetization directly
but serve as a finite electron reservoir. The influence of this electron
reservoir on the magnetization oscillations has been studied recently \cite
{IMV} only for sharp LLs and in the main part numerically. We shall consider
arbitrary shape of LLs.

Since the total particle number remains constant, this intermediate
situation can be considered as the case $N=const$ with some additional
density of states (DoS). This reservoir part of the DoS does not oscillate
with magnetic field and changes on the scale of Fermi energy. Since only
several LLs near FS are important for the dHvA oscillations and the number
of occupied LLs in organic metals $n_{F}\gg 1$ this reservoir part of DoS
may be considered as constant. For the same reason we assume the LL
degeneracy be equal for all LLs. Then the total density of electron states
may be written in the form 
\begin{equation}
\rho (\varepsilon ,B)=g(B)\sum_{n=0}^{\infty }D_{n}(\varepsilon -\hbar
\omega _{c}(n+\frac{1}{2}))\ ,  \label{ro}
\end{equation}
where the LL degeneracy $g(B)=2B/\Phi _{0}+n_{R}\cdot \hbar \omega _{c}$
includes two spin polarizations and the reservoir density of states $%
n_{R}=const$, $\Phi _{0}=2\pi \hbar c/e$ is the flux quantum. $%
D_{n}(\varepsilon )$ is normalized to unity and gives the shape of the n-th
LL with some spin-splitting and the reservoir part of the DoS.

To calculate the magnetization we need first to calculate the chemical
potential. It is given by the equation 
\begin{equation}
N=\int \frac{\rho (\varepsilon ,B)d\varepsilon }{1+\exp \left( \frac{%
\varepsilon -\mu }{T}\right) }  \label{N1}
\end{equation}
It is convenient to count the energy and the chemical potential from the
last occupied LL: 
\begin{equation}
\delta \mu \equiv \mu -\hbar \omega _{c}(n_{F}+1/2)\ ;\ E\equiv \varepsilon
-\hbar \omega _{c}(n_{F}+1/2)  \label{mu1}
\end{equation}
where $n_{F}=$Int$[E_{F}/\hbar \omega _{c}]$ is the number of completely
filled LLs when $\delta \mu =0$\ . It is an integer and jumps by unity one
time on each dHvA period.

Now we apply the direct summation over the LLs. Another approach of the
harmonic expansion of the DoS (\ref{ro}) will be used in the next section
for the case $\mu =const$. At $T\ll \hbar \omega _{c}$ only three LLs near
FS produce the temperature dependent contribution to the oscillations
because the contribution from other LLs is small by a factor $\exp
^{n}(-\hbar \omega _{c}/T)$. The equation (\ref{N1}) then simplifies 
\begin{equation}
\tilde{n}+1=\int_{-3\hbar \omega _{c}/2}^{3\hbar \omega _{c}/2}\frac{D(E)\ dE%
}{1+\exp \left( \frac{E-\delta \mu }{T}\right) }  \label{eq1}
\end{equation}
where $\tilde{n}=N/g-n_{F}=F/B-n_{F}$ is the filling factor of the last
occupied LL, N is the area electron density, $F=const$ is the dHvA
frequency. $\tilde{n}(B)$ is an oscillating function of the magnetic field.
On each dHvA period it monotonically decreases from $1$ to $0$ with
increasing of the magnetic field. The DoS function $D(E)$ here is periodic $%
D(E+\hbar \omega _{c})=D(E)$ and normalized to unity 
\begin{equation}
\int_{-\hbar \omega _{c}/2}^{\hbar \omega _{c}/2}D(E)dE=1  \label{norm}
\end{equation}
The LLs just above and just below the last occupied LL are important when
the chemical potential is situated just between two LLs ($|\delta \mu
|-\hbar \omega _{c}/2\sim T$). Then these two LLs are in equal position and
hence have the same shape. Therefore, we have written the same DoS function $%
D(E)$ for all three LLs.

Now we have to calculate the thermodynamic potential. By definition 
\begin{equation}
\Omega =-k_{B}T\int d\varepsilon \rho (\varepsilon ,B)\ln \left( 1+\exp 
\frac{\mu -\varepsilon }{k_{B}T}\right) \ ,  \label{1}
\end{equation}
where the density of states is given by (\ref{ro}). As in the derivation of (%
\ref{eq1}) we calculate the temperature dependent contribution only from
three LLs near FS. Actually we could leave only one LL but now the
derivation is more general. The expression for the thermodynamic potential
now becomes 
\[
\Omega =-Tg\sum_{n=0}^{n_{F}-2}\frac{\mu -\hbar \omega _{c}(n+1/2)}{T}- 
\]
\[
-Tg\int_{-3\hbar \omega _{c}/2}^{3\hbar \omega _{c}/2}dED(E)\ln \left(
1+\exp \frac{\delta \mu -E}{T}\right) 
\]
The summation in the first term is easy and gives $g\hbar \omega
_{c}(n_{F}-1)^{2}/2-g\mu (n_{F}-1)$. We need the free energy 
\[
F=\Omega +\mu N=g\mu (\tilde{n}+1)+g\hbar \omega _{c}\frac{(n_{F}-1)^{2}}{2}%
- 
\]
\begin{equation}
-Tg\int_{-3\frac{\hbar \omega _{c}}{2}}^{3\frac{\hbar \omega _{c}}{2}%
}dED(E)\ln \left( 1+\exp \frac{\delta \mu -E}{T}\right)  \label{fe}
\end{equation}
To obtain the magnetization $M=-dF/dB$ one should differentiate all the
magnetic field dependent quantities in this expression. As a result one
obtains a quite huge formula. We do not write down it here because it is
useless for the analytical calculations. Even for numerical calculations it
is simpler to use (\ref{fe}). The elegant expression for the magnetization
can be obtained if many LLs are occupied ($n_{F}\gg 1$). This condition is
always satisfied in organic metals where $n_{F}\sim 300$. Then in the third
term of (\ref{fe}) one can differentiate only oscillating quantities since
they give additional factor $n_{F}$. The only such a quantity is $\delta \mu 
$. We assume the function $D(E)$ to be independent of the position of the
chemical potential, that is usually well satisfied. The terms with $\frac{%
d\delta \mu }{dB}$ in $dF/dB$ cancel each other after using (\ref{mu1}) and (%
\ref{eq1}). As a result we get 
\begin{equation}
M\approx C\left[ \frac{1}{2}-\tilde{n}+\frac{\delta \mu }{\hbar \omega _{c}}%
\right]  \label{Mf1}
\end{equation}
where $\delta \mu $ is the solution of (\ref{eq1}). The prefactor 
\begin{equation}
C\equiv \frac{g}{B}\hbar \omega _{c}n_{F}\approx \frac{g}{B}E_{F}=const
\label{C}
\end{equation}
is the same for different dHvA periods since $g\sim B$. The first two terms
in square brackets in (\ref{Mf1}) give the sawtooth form of magnetization,
and the last term $\frac{\delta \mu }{\hbar \omega _{c}}$ determines the
damping of the oscillations due to finite temperature and the LL broadening.
The expression in square brackets of (\ref{Mf1}) is the oscillating part of
the chemical potential divided by $\hbar \omega _{c}$. This can be easily
shown using the definition (\ref{mu1}) of $\delta \mu $. Then the formula (%
\ref{Mf1}) becomes 
\begin{equation}
M=C(\mu -E_{F})/\hbar \omega _{c}  \label{M1mu}
\end{equation}
Another (more apparent) derivation of this relation between the oscillations
of the magnetization and the chemical potential (formula (\ref{Mmu}) ) and
its discussion will be given in section IV.

So, in this section we have derived the formulas (\ref{eq1}) for the
chemical potential and (\ref{Mf1}) for magnetization at finite temperature
and arbitrary DoS distribution. The density of electron reservoir states was
assumed to be approximately constant on the scale of $\hbar \omega _{c}.$ In
the next two sections we shall use these formulas to obtain more concrete
results that are useful for the analysis of the dHvA effect.

\section{Relation between the magnetization and the DoS function}

Now we derive the relation between the wave form of the the magnetization $%
M(B)$ and the DoS distribution $D(E)$, that allows the direct measurement of
the function $D(E)$ at different magnetic field values and spin-splitting if
the temperature smearing is much less than the LL broadening. We assume $%
T\ll \Gamma $, where $\Gamma $ is the width of LLs. Since we need the
function $M(B)$ on one particular dHvA period ($0<\tilde{n}<1$) it is
convenient to consider magnetization as a function of $\tilde{n}\approx
n_{F}(B_{0}-B)/B_{0}$, where on this period $\tilde{n}(B_{0})=0$.
Differentiating (\ref{Mf1}) we obtain 
\begin{equation}
\frac{dM}{d\tilde{n}}=C\left[ -1+\frac{1}{\hbar \omega _{c}}\frac{d\delta
\mu }{d\tilde{n}}\right]  \label{dM}
\end{equation}

Differentiating the equation (\ref{eq1}) with respect to $\delta \mu $ we
get 
\begin{equation}
\frac{d\tilde{n}}{d\delta \mu }=\int_{-3\hbar \omega _{c}/2}^{3\hbar \omega
_{c}/2}\frac{D(E)\ dE}{4T\cosh ^{2}\left( \frac{E-\delta \mu }{2T}\right) }
\label{eq2}
\end{equation}
At $T\ll \Gamma $ the important region of the integration is $|E-\delta \mu
|\sim T$ and the function $D(E)$ can be expanded in Tailor series near this
point 
\begin{equation}
\frac{d\tilde{n}}{d\delta \mu }=\int_{-\frac{3\hbar \omega _{c}}{2}}^{\frac{%
3\hbar \omega _{c}}{2}}\frac{D(\delta \mu )+D^{\prime \prime }(\delta \mu )%
\frac{(E-\delta \mu )^{2}}{2}+..}{4T\cosh ^{2}\left( \frac{E-\delta \mu }{2T}%
\right) }dE  \label{Tailor}
\end{equation}
The all odd terms drop out because the integrand should be a symmetric
function of $(E-\delta \mu )$. We shall keep only the $T^{2}$ term. The
integration can be extended to infinity. After the integration we get 
\begin{equation}
\frac{d\tilde{n}}{d\delta \mu }=D(\delta \mu )+T^{2}D^{\prime \prime
}(\delta \mu )\frac{\pi ^{2}}{6}+O(T^{4})  \label{eq4}
\end{equation}
Substituting this into (\ref{dM}) we obtain 
\begin{equation}
\frac{dM}{d\tilde{n}}=C\left[ -1+\frac{1}{\hbar \omega _{c}\left( D(\delta
\mu )+T^{2}D^{\prime \prime }(\delta \mu )\frac{\pi ^{2}}{6}+O(T^{4})\right) 
}\right]  \label{dM1}
\end{equation}
The temperature dependent terms in (\ref{dM1}) are small and can be
separated by taking the measurements at several low temperatures and
extrapolating to $T=0$. So, one can measure the function 
\begin{equation}
\frac{dM_{T=0}(\tilde{n})}{d\tilde{n}}=C\left[ -1+\frac{1}{\hbar \omega
_{c}D(\delta \mu )}\right]  \label{dM0}
\end{equation}

On the experiment it is very difficult to obtain the proportionality
coefficient between the measured signal and the magnetization. For this
reason we shall look at the constant $C$ as on an unknown factor and later
we shall describe how one can obtain it. Now suppose one can measure the
ratio 
\begin{equation}
R(\tilde{n})\equiv \frac{M_{T=0}(\tilde{n})}{C}\approx \left[ \frac{1}{2}-%
\tilde{n}+\frac{\delta \mu }{\hbar \omega _{c}}\right]  \label{R}
\end{equation}
With this function one can rewrite the equation (\ref{dM0}) as 
\begin{equation}
D(\delta \mu (\tilde{n}))=\frac{1}{\hbar \omega _{c}\left( \frac{dR(\tilde{n}%
)}{d\tilde{n}}+1\right) }  \label{Rm1}
\end{equation}
The function $D(\delta \mu (\tilde{n}))$ is not $D(\delta \mu )=D(E)$. The
dependence $\delta \mu (\tilde{n})$ is simply related to the function $R(%
\tilde{n})$ since the formula (\ref{R}) can be casted into 
\begin{equation}
\delta \mu (\tilde{n})=\hbar \omega _{c}\left[ \tilde{n}-\frac{1}{2}+R(%
\tilde{n})\right]  \label{mur}
\end{equation}
The DoS distribution $D(E)$ is just the plot of $D(\tilde{n})$ (\ref{Rm1})
as a function of $\delta \mu (\tilde{n})$ (\ref{mur}).

Now we have to say how one can get the constant $C$ by which the measured
torque should be divided to obtain the function $R(\tilde{n})$ (\ref{R}).
The normalization condition (\ref{norm}) does not give this constant since
for any C the described procedure gives automatically normalized DoS
distribution. The constant $C$ determines the strength of the magnetization
and, hence, the DoS oscillations. The smaller one takes the constant C in (%
\ref{R}) the larger oscillations of the DoS function he obtains. There is
some critical value $C_{0}$ such that if one assumes $C<C_{0}$ he gets a
singularity in the DoS, calculated from (\ref{Rm1}). At $C=C_{0}$ this
singularity is in the middle of a LL. The peak value $D(0)$ of the DoS is
indeed large compared to the average DoS $\bar{D}=1/\hbar \omega _{c}$ for
small LL broadening $\Gamma \ll \hbar \omega _{c}$. The correct value of the
constant $C$ is then close to $C_{0}$ which gives the function $D(E)$
accurately always except the vicinity of the peaks of LLs. We shall use this
singularity to obtain more accurate value of the constant $C$.

The derivative (\ref{dM0}) has a sequence of the periodically situated
minima that occur on each dHvA period when the chemical potential crosses
the position in the middle of a LL ($\delta\mu =0$). Since $n_F \gg 1$ and
the magnetization is measured on many dHvA periods these minima form a
smooth function of magnetic field 
\[
M^{\prime}_{min}(B) = C\left[ -1 + \frac{1 }{\hbar\omega_c D(0) } \right] 
\]
This function monotonically decreases to some finite limit since $%
\hbar\omega_c D(0)$ increases with the increasing of the magnetic field.
This is because $\hbar\omega_c \sim B$ while the peak value of the DoS $D(0)$
has the very slow dependence on the magnetic field. In the vicinity of the
peaks the DoS distribution may be given by only two parameters: the width of
the LL $\Gamma $ and the peak value $D(0)=1/\alpha\Gamma $. The width $%
\Gamma $ is determined mainly by the impurity scattering which is
approximately independent of the magnetic field. The parameter $\alpha \sim
1 $ depends on the DoS distribution but also is almost independent of the
magnetic field. For Lorentzian shape of LLs $\alpha =\pi $.

Hence, one can assume $D(0)=1/\alpha\Gamma =const$ and from the curve of the
minimum values of 
\begin{equation}
M^{\prime}_{min}(B)=C\left[ -1 + \frac{\alpha\Gamma }{\hbar\omega_c } \right]
\label{M-}
\end{equation}
one can easily obtain two unknown constants: $\alpha\Gamma$ and $C$. This
value of the constant $C$ is accurate enough to calculate $D(E)$ even in the
vicinity of the maxima.

The above procedure gives the DoS at the position of the chemical potential
while the magnetic field and the chemical potential are varied. In general,
this is not the same as if one fix the magnetic field and the chemical
potential but vary the energy itself. In our analysis we disregard the
dependence of $D(E)$ on the position of the chemical potential and hence
this difference. This approximation works when the DoS distribution is
determined by one-electron processes. These are electron scattering on
lattice imperfections and inhomogeneities, the finite probability of the
interlayer jumping and so on. When the many particle effects play an
important role (for example, the change of the magnetic field drives a
sequence of phase transitions as in fractional quantum Hall effect), our
results are not applicable. We have neglected the nontrivial many body
effects at the beginning when have written the formulas (\ref{N1}) and (\ref
{1}) assuming that the system may be described by the distribution of the
single fermion states (\ref{ro}). Nevertheless, this has a very wide
application region because usually fine many particle effects are damped by
impurities, finite $k_{Z}$ dispersion and other factors, especially when
many LLs are occupied.

One can not get the value of LL separation $\hbar \omega _{c}$ without
performing some more measurements because after we have neglected the
temperature dependent terms, the LL separation becomes the only parameter
with the dimensionality of energy. To find it one needs to compare it with
some other energy parameter. If one knows the g-factor of the spin-splitting
in this compound he can compare $\hbar \omega _{c}$ with the spin-splitting
energy. One can also use the temperature dependence of the amplitude of
magnetization oscillations to obtain $\hbar \omega _{c}$ , but this would be
less accurate than the cyclotron resonance method. So, there are many ways
to get $\hbar \omega _{c}$.

To summarize, the suggested procedure of extracting of the DoS distribution
from the oscillations of magnetization consists of the two steps. The first
is to obtain the constant $C$ (that normalizes the measured signal) and the
second to plot $D(E)$ itself. To do this one should measure the torque as a
function of $x=F/B$ where $F=const$ and $x$ changes by unity on each dHvA
period. Then $x=n_F+\tilde n$. This function $M(x)$ should be measured at
several low temperatures and extrapolated to $T=0$. As a result one gets an
oscillating function $M_0(x)$ with period equal to unity. We also need its
derivative $M_0^{\prime}(x)=dM_0(x)/dx$ which is also an oscillating
function. The minima of $M_0^{\prime}(x)$ form a smooth monotonic function $%
M_{min}(x)$. Extrapolating this function to $x=0$ one gets the constant 
\[
C=\lim_{x\rightarrow 0} M_{min}(x) 
\]
Or one can use the formula (\ref{M-}) and obtain the constant $C$ from $%
M_{min}(x)$ more accurately. Substituting the functions $R(x)=M_0(x)/C$ and $%
R^{\prime}(x)=M_0^{\prime}(x)/C$ into (\ref{Rm1}) and (\ref{mur}) one can
plot the expression (\ref{Rm1}) as a function of (\ref{mur}) on one period
of the oscillations. This plot is the desired DoS distribution at the
magnetic field corresponding to the chosen dHvA period.

The described procedure allows to measure the DoS distribution on the Fermi
level for different magnetic field values (because one can obtain $D(E)$ on
each dHvA period) and for different spin-splitting energy (because one can
tilt the magnetic field with respect to the conducting plane of a sample).
This information about the DoS distribution is very important for the study
the role of different scattering mechanisms in the electron motion at
different external parameters. Such an information can not be obtained using
the previous methods of processing the data of the dHvA measurements where
one assumed some particular shape $D(E)$ and then numerically calculated the
magnetization. The results of the calculation (usually only of the envelope)
one compared with the measured signal and if the agreement was good enough
the chosen function $D(E)$ was taken as a result. Our method is more simple
and accurate.

\section{The envelope of magnetization oscillations}

We can also calculate the envelope of magnetization oscillations for several
simple DoS distributions in the limit $\hbar \omega _{c}\gg \Gamma \gg T$.
At low temperature the envelope turns out to depend strongly on the shape of
the LLs while the additional constant contribution to the DoS from other
parts of FS almost does not change the envelope. The envelope of
magnetization oscillations does not give as much information about
electronic structure of the compounds as the DoS distribution does.
Nevertheless it is still useful for the analysis of the dHvA effect and is
quite easy for measurement.

To calculate the envelope of magnetization oscillations we shall take the
total density of states function in the form

\begin{equation}
D(E)=(1-\kappa )\frac{D_{0}\left( E/\Gamma \right) }{\Gamma }+\frac{\kappa }{%
\hbar \omega _{c}}  \label{Dk}
\end{equation}
where $\Gamma $ is the width of LLs, $\kappa <0$ is a number that determines
the constant part of the DoS and 
\[
\int_{-\hbar \omega _{c}/2}^{\hbar \omega _{c}/2}D_{0}\left( \frac{E}{\Gamma 
}\right) \frac{dE}{\Gamma }=1 
\]
converges rapidly. Then the function $D(E)$ is also normalized to unity. The
degeneracy $g(B)$ of LLs should be renormalized to include all additional
parts of FS.

How does the envelope of magnetization depend on $\kappa $. The answer is
that in the limit $n_F \gg 1$ the constant part of DoS affects only the
shape of magnetization but not the envelope of the oscillations. This is
different from the case $\mu =const$ where the constant part of DoS does not
change both the envelope and the shape of the oscillations.

To show this we first should substitute (\ref{Dk}) into (\ref{eq1}). The LLs
just above and just below the last occupied one contribute only when $%
|\delta \mu |\approx \hbar \omega _{c}/2$. As we shall see, if $T,\Gamma \ll
\hbar \omega _{c}$ the extrema of magnetization take place when $|\delta \mu
|\ll \hbar \omega _{c}/2$. So, the regions where $(\hbar \omega
_{c}/2-|\delta \mu |)\sim T\ll \hbar \omega _{c}/2$ are not important for
the envelope of magnetization and $T$ and $\Gamma $ dependent contribution
is given by only one LL. Then the equation (\ref{eq1}) becomes 
\begin{equation}
\tilde{n}=\int_{-\hbar \omega _{c}/2}^{\hbar \omega _{c}/2}\frac{D(E)\ dE}{%
1+\exp \left( \frac{E-\delta \mu }{T}\right) }  \label{seq}
\end{equation}
The integral 
\[
\int_{-\hbar \omega _{c}/2}^{\hbar \omega _{c}/2}\frac{dE}{1+\exp \left( 
\frac{E-\delta \mu }{T}\right) }\approx \frac{\hbar \omega _{c}}{2}+\delta
\mu 
\]
and after substitution of (\ref{Dk}) the equation (\ref{seq}) acquires the
form 
\begin{equation}
\tilde{n}=\int_{-\hbar \omega _{c}/2}^{\hbar \omega _{c}/2}\frac{(1-\kappa
)D_{0}(E)\ dE}{1+\exp \left( \frac{E-\delta \mu }{T}\right) }+\kappa \left( 
\frac{1}{2}+\frac{\delta \mu }{\hbar \omega _{c}}\right)  \label{eq6}
\end{equation}
Substituting this into the expression (\ref{Mf1}) for the magnetization we
get 
\begin{equation}
M=\frac{g\ast }{B}\hbar \omega _{c}n_{F}\left[ \frac{1}{2}+\frac{\delta \mu 
}{\hbar \omega _{c}}-\tilde{n}_{0}\right]  \label{MK}
\end{equation}
where $g\ast =(1-\kappa )g$ is the LL degeneracy of only the oscillating
part of DoS and 
\begin{equation}
\tilde{n}_{0}(\delta \mu )=\int_{-\hbar \omega _{c}/2}^{\hbar \omega _{c}/2}%
\frac{D_{0}(E)\ dE}{1+\exp \left( \frac{E-\delta \mu }{T}\right) }
\label{n0}
\end{equation}
The magnetization on each dHvA period may be considered as a function of $%
\delta \mu :\,M(B)=M(\delta \mu )$. The envelope of the magnetization
oscillations is then given by 
\begin{equation}
M_{\pm }(B)=M(\delta \mu _{ex}(B))=C^{\ast }\left[ \frac{1}{2}+\frac{\delta
\mu _{ex}}{\hbar \omega _{c}}-\tilde{n}_{0}(\delta \mu _{ex})\right]
\label{Men}
\end{equation}
where $C^{\ast }=$ $g^{\ast }\hbar \omega _{c}n_{F}/B$\ and the extremum
values $\delta \mu _{ex}(B)$ at which the magnetization has maxima or minima
are given by the equation 
\begin{equation}
\frac{dM}{dB}=0\,\Leftrightarrow \,\frac{dM(\delta \mu )}{d\delta \mu }%
=C^{\ast }\left( \frac{1}{\hbar \omega _{c}}-\frac{d\tilde{n}}{d\delta \mu }%
\right) =0  \label{ex}
\end{equation}
After accounting for (\ref{eq6}) this equation becomes 
\[
\frac{d\tilde{n}}{d\delta \mu }=\int_{-\hbar \omega _{c}/2}^{\hbar \omega
_{c}/2}\frac{(1-\kappa )D_{0}(E)\ dE}{4T\cosh ^{2}\left( \frac{E-\delta \mu
_{ex}}{2T}\right) }+\frac{\kappa }{\hbar \omega _{c}}=\frac{1}{\hbar \omega
_{c}}\ \Rightarrow 
\]
\begin{equation}
\int_{-\hbar \omega _{c}/2}^{\hbar \omega _{c}/2}\frac{D_{0}(E)\ dE}{4T\cosh
^{2}\left( \frac{E-\delta \mu _{ex}}{2T}\right) }=\frac{1}{\hbar \omega _{c}}
\label{eq7}
\end{equation}
The equation (\ref{eq7}) for $\delta \mu _{ex}$ is independent of $\kappa $
and so does the envelope of magnetization given by (\ref{Men}).

The function $\delta \mu (B)$ is monotonic on each dHvA period and different
on different dHvA periods. Moreover, if $D(E)$ is a symmetric function then $%
\delta \mu $ is an antisymmetric function of $(\tilde{n}-1/2)$ that can be
obtained from (\ref{eq1}) using the identity 
\[
\frac{1}{1+\exp \frac{E-\delta \mu }{T}}=\frac{1}{2}-\frac{1}{2}\frac{\sinh 
\frac{E}{T}-\sinh \frac{\delta \mu }{T}}{\cosh \frac{E}{T}+\cosh \frac{%
\delta \mu }{T}} 
\]
The magnetization (\ref{Mf1}) then is also an antisymmetric function of $(%
\tilde{n}-1/2)$.

In the limit $T\ll \Gamma $ the function $D_{0}(E)$ may be expanded in
Tailor series as has been done in (\ref{Tailor}). If we rest only $\sim
T^{2} $ terms the equation (\ref{eq7}) becomes

\begin{equation}
D_0(\delta\mu_{ex} )+ T^2 D_0^{\prime\prime}(\delta\mu_{ex} ) \frac{\pi ^2}{6%
} =\frac{1}{\hbar\omega_c}  \label{eqt}
\end{equation}
Formula(\ref{n0}) can also be simplified. Integrating by parts and expanding
up to the terms $\sim T^2$ we get

\begin{equation}
\tilde{n}_{0}=G_{0}(\delta \mu )+T^{2}D_{0}^{\prime }(\delta \mu )\frac{\pi
^{2}}{6}  \label{Gn}
\end{equation}
where we have introduced the function 
\begin{equation}
G_{0}(E)=\int_{-\hbar \omega _{c}/2}^{E}D_{0}(E^{\prime })dE^{\prime }
\label{GD}
\end{equation}
The function $G_{0}(E)$ is dimensionless and changes in the range $[0;1]$.
From the equation (\ref{eq7}) one can immediately say that if $D_{0}(x)$ has
the exponentially falling tails, the $\delta \mu _{ex}\sim \pm \Gamma \ln
(\hbar \omega _{c}/\Gamma )$. If $D_{0}(x)$ falls off as $x^{-\gamma }$,
then $\delta \mu _{ex}\sim \Gamma (\hbar \omega _{c}/\Gamma )^{1/\gamma }$.
This determines the behavior of the envelope (\ref{Men}). Now we shall
consider in details two different functions $D_{0}(x)$.

\subsection{ Exponentially falling $D(x)$}

There are many symmetric exponentially falling functions that are eligible
to be a density of states function $D_0(E)$. They all lead to the close
expressions for the envelope of the magnetization oscillations and for
example we shall take 
\begin{equation}
D_0(E) = \frac{1}{4\Gamma \cosh ^2 \left( \frac{E}{2\Gamma } \right) }
\label{Dex}
\end{equation}
Now the equation (\ref{eq7}) possesses the symmetry $T \leftrightarrow
\Gamma $ if $T$ or $\Gamma \ll \hbar\omega_c$ and one can easily obtain the
limit $\Gamma \ll T$ from the limit $T\ll\Gamma $.

The equation (\ref{eqt}) can be solved by the iteration procedure with small
parameter $T^2/\Gamma^2$. In zeroth approximation 
\[
D_0(\delta\mu^0 _{ex} )= \frac{1}{4\Gamma \cosh ^2 \left( \frac{%
\delta\mu_{ex}^0}{2\Gamma } \right) } = \frac{1}{\hbar\omega_c} 
\]
Since the magnetization (\ref{Mf1}) is an antisymmetric function of $%
\delta\mu $ we shall consider only one (negative) root of this equation 
\begin{equation}
\delta\mu_{ex}^0 = -2\Gamma arccosh \sqrt{\frac{\hbar\omega_c}{4\Gamma } }
\approx -\Gamma \ln \left( \frac{\hbar\omega_c}{\Gamma } \right)  \label{mu0}
\end{equation}

From equation (\ref{Gn}) we get 
\begin{equation}
\tilde{n}_{0}(\delta \mu _{ex}^{0})\approx \frac{1}{1+\frac{\hbar \omega _{c}%
}{\Gamma }}  \label{ne0}
\end{equation}
Substituting (\ref{mu0}) and (\ref{ne0}) into (\ref{Men}) we get the
envelope in zeroth approximation 
\begin{equation}
M_{\pm }^{0}=\pm C^{\ast }\left[ \frac{1}{2}-\frac{\Gamma }{\hbar \omega _{c}%
}\ln \left( \frac{\hbar \omega _{c}}{\Gamma }\right) -\frac{\Gamma }{\hbar
\omega _{c}}\right]  \label{M0ex}
\end{equation}

If one makes the replacement $\Gamma \rightarrow T$ this result coincides
with the result of Vagner et al. \cite{V1} , obtained for the case $\Gamma
=0 $ and finite temperature. It's not surprising because of the mentioned
symmetry $\Gamma \leftrightarrow T$ with the DoS function (\ref{Dex}).

One can easily obtain the first temperature correction to the envelope (\ref
{M0ex}). In the first approximation the equation (\ref{eqt}) becomes 
\begin{equation}
D_{0}(\delta \mu _{ex})+T^{2}D_{0}^{\prime \prime }(\delta \mu _{ex}^{0})%
\frac{\pi ^{2}}{6}=\frac{1}{\hbar \omega _{c}}  \label{eqt1}
\end{equation}
The correction to the chemical potential 
\begin{equation}
\Delta _{T}\mu _{ex}\equiv \delta \mu _{ex}^{1}-\delta \mu _{ex}^{0}=\frac{%
T^{2}D_{0}^{\prime \prime }(\delta \mu _{ex}^{0})\frac{\pi ^{2}}{6}}{%
D_{0}^{\prime }(\delta \mu _{ex}^{0})}  \label{dmu1}
\end{equation}
This correction $\Delta _{T}\mu \sim T^{2}$. The first correction to the
magnetization $\Delta M_{+}\sim (\Delta _{T}\mu )^{2}$ because according to
the equation \ref{ex} $M^{\prime }(\delta \mu _{ex})=0$ and the correction
to the envelope of magnetization 
\[
\Delta M_{+}=M(\delta \mu _{ex}^{1})-M(\delta \mu _{ex}^{0})=M^{\prime
}(\delta \mu _{ex}^{0})\,\Delta _{T}\mu _{ex}+\frac{d^{2}M(\delta \mu
_{ex}^{0})}{d(\delta \mu _{ex}^{0})^{2}}\frac{(\Delta _{T}\mu _{ex})^{2}}{2}%
= 
\]
\begin{equation}
=-\frac{d^{2}M(\delta \mu _{ex}^{0})}{d(\delta \mu _{ex}^{0})^{2}}\frac{%
(\Delta _{T}\mu _{ex})^{2}}{2}=C^{\ast }\frac{d^{2}\tilde{n}_{0}(\delta \mu
_{ex}^{0})}{d(\delta \mu _{ex}^{0})^{2}}\frac{(\Delta _{T}\mu _{ex})^{2}}{2}
\label{dmt}
\end{equation}
In writing this we have used $M^{\prime }(\delta \mu _{ex}^{0})=M^{\prime
}(\delta \mu _{ex})-$\ $M^{\prime \prime }(\delta \mu _{ex}^{0})\Delta
_{T}\mu _{ex}$. From (\ref{Gn}) we have 
\begin{equation}
\tilde{n}_{0}^{\prime \prime }(\delta \mu _{ex}^{0})\approx D_{0}^{\prime
}(\delta \mu _{ex}^{0})  \label{tiln}
\end{equation}
In our case (\ref{Dex}) 
\[
D_{0}^{\prime }(\delta \mu _{ex}^{0})=-\frac{\sinh \left( \frac{\delta \mu
_{ex}^{0}}{2\Gamma }\right) }{4\Gamma ^{2}\cosh \left( \frac{\delta \mu
_{ex}^{0}}{2\Gamma }\right) }\approx \frac{1}{\Gamma \hbar \omega _{c}}\ , 
\]
\[
D_{0}^{\prime \prime }(\delta \mu _{ex}^{0})\approx \frac{3}{2\Gamma
^{2}\hbar \omega _{c}}\ \mbox{ and }\ \Delta _{T}\mu _{ex}=-\frac{T^{2}}{%
\Gamma }\frac{\pi ^{2}}{4} 
\]
Substituting this into (\ref{dmt}) and (\ref{tiln}) and using (\ref{M0ex})
we obtain the envelope in the first approximation 
\begin{equation}
M_{\pm }^{1}=C^{\ast }\left[ \frac{1}{2}-\frac{\Gamma }{\hbar \omega _{c}}%
\ln \left( \frac{\hbar \omega _{c}}{\Gamma }\right) -\frac{\Gamma }{\hbar
\omega _{c}}-\frac{\Gamma }{\hbar \omega _{c}}\frac{1}{2}\left( \frac{\pi
^{2}}{4}\right) ^{2}\left( \frac{T}{\Gamma }\right) ^{4}\right]  \label{M+ex}
\end{equation}

\subsection{ Lorentzian shape of LLs}

For the Lorentzian shape of LLs 
\begin{equation}
D_{0}(E)=\frac{1/\pi \Gamma }{1+(E/\Gamma )^{2}}  \label{Dl}
\end{equation}
we shall do the same steps as for the exponentially falling $D(E)$. From
equation (\ref{eqt}) in zeroth approximation we obtain 
\begin{equation}
\delta \mu _{ex}^{0}=\Gamma \sqrt{\frac{\hbar \omega _{c}}{\pi \Gamma }-1}%
\approx \sqrt{\Gamma \hbar \omega _{c}/\pi }  \label{mu0l}
\end{equation}
and 
\begin{equation}
\tilde{n}_{0}(\delta \mu _{ex}^{0})=\frac{1}{\pi }\arctan \left( \frac{%
\delta \mu _{ex}^{0}}{\Gamma }\right) +\frac{1}{2}  \label{n0l}
\end{equation}
Substituting this into the expression for magnetization (\ref{Men}) we get 
\begin{equation}
M_{\pm }^{0}=\pm C^{\ast }\left[ \frac{1}{2}-2\sqrt{\frac{\Gamma }{\pi \hbar
\omega _{c}}}\right]  \label{M0l}
\end{equation}
This is different from (\ref{M0ex}). So, the envelope is different for
different shapes of LLs. The temperature correction is also different.

To obtain the first temperature correction we need

\[
D_{0}^{\prime }(\delta \mu _{ex}^{0})\approx \frac{2}{\hbar \omega _{c}}%
\sqrt{\frac{\pi }{\Gamma \hbar \omega _{c}}}\mbox{ and }D_{0}^{\prime \prime
}(\delta \mu _{ex}^{0})\approx \frac{6\pi }{\Gamma (\hbar \omega _{c})^{2}} 
\]
\[
\Rightarrow \Delta _{T}\mu _{ex}=-\frac{\pi ^{5/2}T^{2}}{2\sqrt{\Gamma \hbar
\omega _{c}}} 
\]
Substituting this into (\ref{dmt}) and (\ref{tiln}) and using (\ref{M0l}) we
get 
\begin{equation}
M_{\pm }^{1}=C^{\ast }\left[ \frac{1}{2}-2\sqrt{\frac{\Gamma }{\pi \hbar
\omega _{c}}}-2\sqrt{\frac{\Gamma }{\pi \hbar \omega _{c}}}\frac{\pi
^{6}T^{4}}{8\Gamma ^{2}(\hbar \omega _{c})^{2}}\right]  \label{M+l}
\end{equation}

\smallskip

The formulas (\ref{M+ex}) and (\ref{M+l}) are not very confident even at low
temperature. This is because the oscillating part of the DoS distribution
function depends not only on the one parameter -- the width $\Gamma $, but
also on the LL separation $\hbar\omega_c$ and hence on the magnetic field.
Any case the approximate functions (\ref{Dex}) and (\ref{Dl}) may be too far
from reality. The exponentially falling $D_0(E)$ is always the incorrect
approximation, while the confidence of the Lorentzian shape of LLs in 2D
metals will be discussed later. The envelope of magnetization oscillations
is very sensitive to the tails of DoS function $D(E)$ at low temperature and
one should use more realistic (probably magnetic field dependent) functions $%
D(E)$ for accurate calculations.

Nevertheless, the formulas (\ref{M+ex}) and (\ref{M+l}) can be used to
indicate the qualitative features. They predict that the first temperature
correction to the envelope is proportional to $T^4$ and that this envelope
depends strongly on the shape of LLs at low temperature.

We can now say also at what conditions the previous results are valid. At $%
N=const$ and finite temperature the only results\cite{V1}, \cite{GV} for
magnetization and its envelope have been obtained for sharp LLs ($%
D(E)=\delta (E)$). These formulas have been derived also in the limit $T\ll
\hbar\omega_c$. If we assume the shape of LLs to be Lorentzian these result
are valid if $\hbar\omega_c\gg T\gg \sqrt{\Gamma \hbar\omega_c}$. This
narrow region may not exist at all, that explains why many experimental data
can not be described by these formulas.

\section{Magnetization oscillations at constant chemical potential}

The case of constant chemical potential may take place in organic metals
when the electron reservoir formed by the one-dimensional parts of FS is
very large. This has been observed recently on $\beta ^{\prime \prime }$%
-(BEDT-TTF)$_{2}$SF$_{5}$CH$_{2}$CF$_{2}$SO$_{3}$ \cite{Fix} by the inverse
sawtooth wave form of the magnetization oscillations.

The case $\mu =const$ may take place also due to giant magnetostriction
(volume change) that cancels the variations of \ chemical potential
associated with the oscillations of the density of states. This has been
first observed in beryllium\cite{Al}. In organic metals the electron density
is not usually large enough to observe the giant magnetostriction. The
restriction $\mu =const$ can be achieved, in principle, by connecting the
sample with some big 3D piece of metal.

The case $\mu = const$ is interesting because it is more favorable to
observe the diamagnetic phase transition in 2D metal and Condon domains \cite
{Con}. It is also interesting to compare the cases $\mu = const$ and $N =
const$.

Although in pure 2D materials at low temperature the dHvA oscillations are
essentially non sinusoidal, the expansion over harmonics is still useful and
the infinite harmonic series for some special densities of states can be
calculated with good accuracy. The 2D harmonic expansion of the
magnetization oscillations has been first obtained by Shoenberg\cite{Sh} in
the case $\mu =const$ using the intuitive phase smearing arguments. In the
special case of the Lorentzian shape of LLs it has been derived recently\cite
{Min} using Green's function. We shall give another simple derivation of
this harmonic expansion for arbitrary shape of LLs. By definition the
thermodynamic potential is given by the same expressions (\ref{1},\ref{ro})
as in the case $N=const$. We assume $D(x)$ to be independent of the number
of a LL, since in dHvA effect only the LLs near FS are important and for
these levels the function $D(x)$ does not differ substantially. Then $\rho
(E)=\rho (E/\hbar \omega _{c})$ is a periodic function. One can take the
spin-splitting into account by introducing $\rho _{s}(E)=\rho (\frac{E}{%
\hbar \omega _{c}}-\Delta /2)+\rho (\frac{E}{\hbar \omega _{c}}+\Delta /2)$,
where $\Delta $ is the energy of spin splitting divided by $\hbar \omega
_{c} $. This is equivalent to introducing $\Omega _{s}(\mu )=\Omega (\frac{%
\mu }{\hbar \omega _{c}}-\Delta /2)+\Omega (\frac{\mu }{\hbar \omega _{c}}%
+\Delta /2)$, since integration in (\ref{1}) is between infinite limits.
Spin will be recovered in the final result by introducing $M_{s}(\mu )=M(%
\frac{\mu }{\hbar \omega _{c}}-\Delta /2)+M(\frac{\mu }{\hbar \omega _{c}}%
+\Delta /2)$ , where $M(\mu )$ is the magnetization without spin.

The density of states (\ref{ro}) can be rewritten as a harmonic series 
\begin{equation}
\rho (E,B)=\frac{g}{\hbar \omega _{c}}\sum_{k=-\infty }^{\infty }\exp \left[
2\pi i\left( \frac{E}{\hbar \omega _{c}}-\frac{1}{2}\right) k\right] A(k)%
\frac{1+\mbox{sign}(E)}{2}  \label{rho}
\end{equation}
where we introduced the Fourier transform of the DoS distribution on each LL 
\begin{equation}
A(k)=\int_{-\frac{\hbar \omega _{c}}{2}}^{\frac{\hbar \omega _{c}}{2}%
}D(E^{\prime })\exp \left( 2\pi i\frac{E^{\prime }k}{\hbar \omega _{c}}%
\right) \mbox{d}E^{\prime }  \label{A}
\end{equation}
and the factor $(1+\mbox{sign}(E))/{2}$ indicates that no electron states
exist for $E<0$. One can omit the term with $k=0$ since it doesn't affect
the magnetic oscillations. Substituting (\ref{rho}) into (\ref{1}) and
integrating two times by parts we get

\begin{equation}
\tilde \Omega =-\frac{g}{\hbar \omega _{c}}\int \mbox{d}E\frac{\partial f}{%
\partial E}\sum_{k\neq 0}\frac{\hbar ^2 \omega_{c}^2}{(2\pi k)^{2}}\exp %
\left[ 2\pi i\left( \frac{E}{\hbar \omega _{c}}-\frac{1}{2}\right) k\right]
A(k)  \label{o}
\end{equation}
Since $\frac{\partial f}{\partial E}\neq 0$ only in a small region near FS,
the factor $(1+\mbox{sign}(E))/{2}$ has been omitted.

Introducing the Furrier transform of $\frac{\partial f}{\partial E}$ 
\begin{equation}
F_{T}(k)\equiv \int \mbox{d}E\left( -\frac{\partial f}{\partial E}\right)
\exp \left( 2\pi i\frac{(E-\mu )k}{\hbar \omega _{c}}\right) =\frac{2\pi
^{2}Tk/\hbar \omega _{c}}{\sinh (2\pi ^{2}Tk/\hbar \omega _{c})}\equiv \frac{%
k\lambda }{\sinh k\lambda }  \label{Fur}
\end{equation}
we get 
\begin{equation}
\tilde{\Omega}=\frac{g\hbar \omega _{c}}{2\pi ^{2}}\sum_{k=1}^{\infty }\frac{%
(-1)^{k}}{k^{2}}\cos \left( \frac{\mu }{\hbar \omega _{c}}2\pi k\right)
A(k)\ F_{T}(k)  \label{oh}
\end{equation}
Differentiating (\ref{oh}) we obtain the harmonic expansion of the
magnetization oscillations 
\begin{equation}
\tilde{M}=-\frac{\partial \tilde{\Omega}}{\partial B}\mid _{\mu =const}=%
\frac{g\mu }{\pi B}\sum_{k=1}^{\infty }\frac{(-1)^{k+1}}{k}\sin \left( \frac{%
\mu }{\hbar \omega _{c}}2\pi k\right) A(k)\ F_{T}(k)  \label{6M}
\end{equation}
This coincides with the Shoenberg's formula\cite{Sh}.

The oscillations of the electron density are given by 
\begin{equation}
\tilde{N}=-\frac{\partial \tilde{\Omega}}{\partial \mu }\mid _{B=const}=%
\frac{g}{\pi }\sum_{k=1}^{\infty }\frac{(-1)^{k}}{k}\sin \left( \frac{\mu }{%
\hbar \omega _{c}}2\pi k\right) A(k)\ F_{T}(k)  \label{tN}
\end{equation}
From (\ref{6M}) and (\ref{tN}) the simple relation follows

\begin{equation}
\tilde M=-\mu\tilde N/B.  \label{MN}
\end{equation}
It is valid only in the limit $\mu \gg \hbar \omega_c $ where the formula (%
\ref{6M}) is valid.

In the case $N=const$ one can obtain the similar relation between
magnetization and the oscillations of the chemical potential. The
magnetization oscillations at $N=const$ 
\[
\tilde{M}=-\frac{d(\tilde{\Omega}+N\mu )}{dB}\mid _{N=const}=-\frac{\partial 
\tilde{\Omega}}{\partial B}\mid _{\mu ,N=const}- 
\]
\[
-\left( \frac{\partial \tilde{\Omega}}{\partial \mu }\mid
_{N,B=const}+N\right) \frac{d\mu }{dB}\mid _{N=const}=-\frac{\partial \tilde{%
\Omega}}{\partial B}\mid _{\mu ,N=const} 
\]
So, the formula (\ref{6M}) is valid also in the case $N=const$ but the
chemical potential becomes an oscillating function of the magnetic field.
The total electron density 
\begin{equation}
N=\int_{-\infty }^{\mu }[n_{0}(E)+n_{R}(E)]dE+\tilde{N}=const  \label{Mum1}
\end{equation}
where $n_{0}(E)=(g/\hbar \omega _{c})\theta (E)$ is the average DoS on the
2D part of FS and $n_{R}(E)$ is the reservoir DoS. Substituting $%
N=\int_{-\infty }^{E_{F}}[n_{0}(E)+n_{R}(E)]dE$ and (\ref{MN}) we get 
\begin{equation}
\tilde{M}=\frac{\mu }{B}\int_{E_{F}}^{\mu }[n_{0}(E)+n_{R}(E)]dE
\label{Mmu1}
\end{equation}
If $n_{R}(E)=const$ this simplifies 
\begin{equation}
\tilde{M}=\frac{\mu }{B}(n_{0}+n_{R})(\mu -E_{F}).  \label{Mmu}
\end{equation}
This expression coincides with the formula (\ref{Mf1}). The variations of
the chemical potential $|\mu -E_{F}|<\hbar \omega _{c}/2$ and the relation (%
\ref{Mmu}) is valid if $n_{R}^{\prime }(E_{F})\hbar \omega _{c}/2\ll
n_{R}(E_{F})$. This is usually fulfilled in the limit $\mu /\hbar \omega
_{c}\gg 1$. The formula (\ref{Mmu}) has been obtained previously only for
sharp LLs in \cite{IMV} and for Lorentzian shape of LLs without reservoir
states in \cite{Min}. We have derived this formula in general case so that
its applicability region becomes clear.

The electron reservoir (the independent of the magnetic field DoS) does not
contribute to the magnetization oscillations at $\mu =const$. This is
evident from (\ref{1}) since $\Omega $ is linear in $\rho (E,B)$ and the DoS 
$\rho (E,B)$ is the only magnetic field dependent factor in right hand of
the expression (\ref{1}). So, in the case $\mu =const$ one can consider only
oscillating part of the DoS.

The expression (\ref{6M}) depends essentially on the function $A(k)$ that is
determined by shape of LLs. The question about the DoS distribution on the
LLs in 2D metals is open. The Lorentzian shape of LLs (\ref{Dl}) at small LL
broadening $\Gamma $ leads to the Dingle low of harmonic damping

\begin{equation}
A(k)=\exp \left( -\frac{\Gamma k}{\hbar \omega _{c}}\right)  \label{A1}
\end{equation}
This exponential low of harmonic damping has a direct physical meaning -- it
assumes the noncoherent scattering of the electrons on the impurities. The
width of LLs $\Gamma $ is related to the electron scattering relaxation time 
$\tau =\pi \hbar /\Gamma $ and (\ref{A1}) is the amplitude of the
probability that an electron returns to the initial state after $k$
cyclotron periods. This can be shown by follows. The DoS is connected with
the electron Green's function by the relation 
\begin{equation}
\rho (E)=-\frac{1}{\pi }ImG^{R}(r,r,E)  \label{RG}
\end{equation}
The imaginary part of the retarded Green's function 
\begin{equation}
ImG^{R}(r,r,E)=\frac{1}{2}\left( G^{R}-G^{A}\right) =-\frac{1}{2}%
\int_{-\infty }^{\infty }i<\psi ^{+}(r,0)\psi (r,t)+\psi (r,t)\psi
^{+}(r,0)>\exp (-iEt/\hbar )dt  \label{IG}
\end{equation}
Substituting this into (\ref{A}) we get 
\[
A(k)=\int_{-\infty }^{\infty }\rho (E)\exp \left( \frac{2\pi iEk}{\hbar
\omega _{c}}\right) dE= 
\]
\[
=\frac{1}{2\pi }\int_{-\infty }^{\infty }dt<\psi ^{+}(r,0)\psi (r,t)+\psi
(r,t)\psi ^{+}(r,0)>\int_{-\infty }^{\infty }dE\exp \left( \frac{2\pi iEk}{%
\hbar \omega _{c}}-\frac{iEt}{\hbar }\right) dE= 
\]
using 
\[
\frac{1}{2\pi }\int_{-\infty }^{\infty }dE\exp \left( \frac{2\pi iEk}{\hbar
\omega _{c}}-\frac{iEt}{\hbar }\right) dE=\delta (t-2\pi k/\omega _{c}) 
\]
we finely obtain 
\begin{equation}
A(k)=<\psi ^{+}(r,0)\psi (r,T_{c}k)+\psi (r,T_{c}k)\psi ^{+}(r,0)>
\label{AR}
\end{equation}
where $T_{c}=2\pi /\omega _{c}$ is the cyclotron period.

The formula (\ref{A1}) states for the arbitrary ratio $\Gamma /\hbar\omega_c$
and is more general than (\ref{Dl}). In 3D metals after each scattering
process an electrons leave the impurity for a long time because it is free
in the direction of the magnetic field. In 2D case the electrons can not
move in $z$ direction and return to the same impurity after each cyclotron
period. If the phase smearing time is longer than the cyclotron period the
electron interfere with itself and this may change the amplitude (\ref{A1})
and hence the shape of LLs. So, the Dingle low of harmonic damping may
violate if the cyclotron energy is greater than warping and the width of LLs 
$\hbar\omega_c \gg W,\Gamma $. Nevertheless on many experiments on organic
metals no remarkable deviation from the Dingle law of harmonic damping has
been observed. This may be because even small $k_{Z}$ dispersion breaks the
coherence of 2D scattering. We now assume the Dingle low (\ref{A1}) to be
valid and try to sum the series (\ref{6M}) for magnetization oscillations at
arbitrary $T$ and $\Gamma $.

The oscillating part of the magnetization is 
\begin{equation}
\tilde{M}=\frac{g\mu }{\pi B}Im\sum_{k=1}^{\infty }\exp \left[ iy k-b\,k%
\right] (-1)^{k+1}\frac{\lambda }{\sinh k\lambda }  \label{11}
\end{equation}

where 
\begin{equation}
\lambda =2\pi ^{2}T/\hbar \omega \ ,\ b=\pi /\omega _{c}\tau \ \mbox{ and }
\ y=\frac{2\pi \mu }{\hbar \omega_c }  \label{lb}
\end{equation}
To calculate this sum we expand 
\begin{equation}
\frac{1}{\sinh (k\lambda )}=\frac{2e^{-k\lambda }}{1-e^{-k\lambda }}%
=\sum_{p=0}^{\infty }2e^{-k\lambda }e^{-2pk\lambda }  \label{12}
\end{equation}

Now sum over k is the sum of geometric series and can be calculated 
\begin{equation}
\tilde{M}=\frac{2g\mu \lambda }{\pi B}Im\sum_{p=0}^{\infty }Q(p)  \label{14}
\end{equation}
where 
\begin{equation}
Q(p)\equiv \frac{\exp \left[ iy-b-\lambda (1+2p)\right] }{1+\exp \left[
iy-b-\lambda (1+2p)\right] }  \label{Q}
\end{equation}

$Q(p)$ is a smooth monotonic function so one can apply the Euler-Maclaurin
summation formula\cite{ZW} 
\begin{equation}
\sum_{p=0}^{\infty }Q(p)=\frac{1}{2}Q(0)+\int_{0}^{\infty }Q(p)\mbox{d}%
p-\sum_{j=1}^{\infty }\frac{Q^{(2j-1)}(0)}{(2j)!}B_{2j}  \label{EM}
\end{equation}
where $B_{k}$ are the Bernoulli numbers. The terms of the sum in the
right-hand part of eq. (\ref{EM}) fall off $\sim \frac{(2\lambda )^{k}}{%
(k+1)!}$ , hence their sum converge rapidly. The first term $\sim \frac{%
\lambda }{12}$ $Q(0)$. So, all the sum in the right-hand part is much less
than $Q(0)$ and we can ignore it. The integral 
\begin{equation}
Im\int_{0}^{\infty }Q(p)dp=\frac{1}{2\lambda }\arctan \left( \frac{\sin (y)\ 
\text{exp}\left[ -b-\lambda \right] }{1+\cos (y)\ \text{exp}\left[
-b-\lambda \right] }\right)  \label{20}
\end{equation}
Substituting (\ref{Q}), (\ref{EM}) and (\ref{20}) into (\ref{14}) and
recovering the spin splitting we obtain 
\[
\tilde{M_{s}}=\frac{g\mu }{\pi B}\left[ \arctan \left( \frac{\sin
(y-\pi\Delta )\exp \left[ -b-\lambda \right] }{1+\cos (y-\pi\Delta )\exp %
\left[ -b-\lambda \right] }\right) +\right. 
\]
\begin{equation}
\left. +\frac{\lambda \sin (y-\pi\Delta ) \exp \left[ -b-\lambda \right] }{%
1+2\cos (y-\pi\Delta )\exp \left[ -b-\lambda \right] +\exp \left[
-2b-2\lambda \right] }\right] +\{(y-\pi\Delta )\rightarrow (y+\pi\Delta )\}
\label{Ms}
\end{equation}

This expression gives the amplitude and the shape of magnetization
oscillations. It is much more convenient for practical use in the limit of
weak harmonic damping than the infinite series (\ref{6M}). In the limit of
strong harmonic damping $\lambda \gg 1$ or $b\gg 1$ formula (\ref{Ms})
correctly gives the first harmonics. In the other limit $\lambda =0 $ the
formula (\ref{Ms}) becomes exact and simple 
\[
\tilde {M_{s}}=\frac{\mu }{\pi \Phi_0}\arctan \left( \frac{ \sin
(y-\pi\Delta ) e^{-b} } {1+\cos (y-\pi\Delta ) e^{-b} }\right) +\{
(y-\pi\Delta )\rightarrow (y+\pi\Delta )\} 
\]
This coincides with the recent result of T. Champel and V.P. Mineev \cite
{Min}. At $b=0$ , using the identity 
\[
\arctan \left( \frac{\sin x}{1+\cos x}\right) =\arctan \left( \frac{\sin
(x/2)}{\cos (x/2)}\right) 
\]
one gets a saw-tooth function as has been predicted by Peierls\cite{Pei}.

\ If spin splitting energy is just n-times the cyclotron energy (this can be
achieved by tilting the magnetic field) and chemical potential $\mu \gg
\hbar \omega _{c}$\ , formula (\ref{Ms}) substantially simplifies 
\begin{equation}
\tilde{M_{s}}=\frac{2g\mu }{\pi B}\left[ \arctan \left( \frac{\sin y\exp %
\left[ -b-\lambda \right] }{1+\cos y\exp \left[ -b-\lambda \right] }\right) +%
\frac{\lambda \sin y\exp \left[ -b-\lambda \right] }{1+2\cos y\exp \left[
-b-\lambda \right] +\exp \left[ -2b-2\lambda \right] }\right]   \label{M1}
\end{equation}
Now\ one can obtain the envelope of magnetization oscillations. The equation 
$\frac{\partial M}{\partial B}=0$ gives 
\begin{equation}
\cos y\ast =\frac{-(1+3ex^{2}+\lambda +\lambda \,ex^{2})+\sqrt{%
(1+3ex^{2}+\lambda +\lambda \,ex^{2})^{2}-8ex^{2}(1+2\lambda +ex^{2})}}{4ex}
\label{cos}
\end{equation}
where $ex\equiv \exp [-b-\lambda ]$, that determines the positions of
extrema of $M(B)$. Substitution this into (\ref{M1}) gives the envelope of
magnetization oscillations. In the limit $\lambda \ll 1,\ b\ll 1,\ \ \lambda
\sim b$ we have 
\begin{equation}
M_{\pm }(B)\simeq \pm \frac{\mu S}{\Phi _{0}}\left[ 1-\frac{2\sqrt{2b}}{\pi }%
+O\left( \lambda ^{\frac{3}{2}},b^{\frac{3}{2}}\right) \right]   \label{M+}
\end{equation}
where $\lambda $ and $b$ are given by (\ref{lb}). \ The result $M\pm $ does
not depend on temperature in first approximation at $\lambda \sim b\ll 1$.
This is not surprising because $A(k)=\exp (-bk)$ has stronger dependence on $%
b$ than $F_{T}(k)$ (eq. \ref{Fur}) on $\lambda $ for small $k$, that are
important in the sum (\ref{6M}). Since $b=2\pi \Gamma /\hbar \omega _{c}$
the formula  (\ref{M+}) gives the same envelope as (\ref{M0l}). Thus, if the
Dingle low of harmonic damping is valid, the envelope of magnetization
oscillations is identical in the cases of constant particle density and
fixed chemical potential in the limit $b=2\pi \Gamma /\hbar \omega _{c}\ll 1$
and zero temperature.

Using formula (\ref{Ms}) one can examine the possibility of the diamagnetic
phase transition(DPT) \cite{Sh}. It must take place if $dM/dB >1/4\pi $. It
has been observed in some pure 3D metals due to the formation of Condon
domains\cite{Con}. In 2D case it can lead to an energy gap at the FS,
analogies to that of QHE. The diamagnetic phase transition is not possible
in available heterostructures since the magnetization is $\sim $5 orders
less than necessary. In organic metals the electron density is much greater
than in heterostructures and the constraint $\mu =const$ is more favorable
to observe the DPT. For example, in the compound $\kappa
-(BEDT-TTF)_{2}I_{3} $ at $T_D=0.1K$ the slope $dM/dB$ turns out to be only
5 times smaller than needed for this phase transition. Maybe there exist
another compound where this derivative $dM/dB$ is great enough.

There is another characteristic parameter of magnetization oscillations,
besides the frequency and the amplitude, that can be compared with the
experiment. It is the value of the magnetic field $B_0$ above which the
spin-splitting is clearly visible on the graph $M(B)$. More formally, it is
the point $B_0$, above which the first derivative of magnetization has an
additional zero on each period due to the spin-splitting. This point $B_0$
is extremely sensitive to the parameters of the system, such as the tilting
angle of the magnetic field, the shape of LLs and the reservoir DoS, the
temperature and so on. But to obtain an accurate theoretical description of
this critical field one needs to make the numerical calculations to take
into account many factors, unimportant for the amplitude of the oscillations.

In the 2D organic metals the magnetic breakdown \cite{S0} usually takes
place at some field $B_{MB}$. The frequency of dHvA oscillations is
different below and above $B_{MB}$ . In the vicinity of $B_{MB}$ one should
apply more complicated theory of the dHvA effect (\cite{F} and the
references therein). But far from $B_{MB}$ all formulas remain valid, but
one should always substitute $\mu $ by $\hbar \omega_c \cdot F/B$ , where $F$
is the new dHvA frequency.

\section{Conclusion}

The dHvA effect in normal 2D metals has been investigated at different
conditions and at the arbitrary shape and width of LLs. Finite temperature,
spin-splitting and electron reservoir states are taken into account. In the
limit $E_F/\hbar\omega_c \gg 1$ the formulas greatly simplifies and many
results can be obtained analytically. In this limit there are simple
relations (\ref{MN}) and (\ref{Mmu}) between the magnetization oscillations
and the oscillations of the chemical potential (in the case $N=const$) or
the oscillations of the electron density (in the case $\mu =const$). These
relations are valid at the arbitrary shape of LLs.

We have obtained the analytical formulas for the magnetization and its
envelope at different conditions and shapes of LLs, that may be useful to
treat the experimental data. The envelope of the magnetization oscillations
depends strongly on the shape of LLs in the limit of weak harmonic damping
(compare the formulas \ref{M+ex} and \ref{M+l}). 

We derived the explicit relation between the wave form of the magnetization
oscillations and the shape of LLs at arbitrary reservoir DoS. This relation
makes possible to obtain the DoS distribution at arbitrary magnetic field
and spin-splitting from the measurement of the dHvA effect. The DoS
distribution gives much more information about the electronic structure of
compounds than the Dingle temperature and the effective electron mass. This
may help to solve the theoretical problem about the DoS distribution in 2D
metals in magnetic field.

\bigskip P.G. thanks Prof. A.M.Dyugaev for useful discussion and Prof.
P.Wyder for the hospitality in Grenoble High Magnetic Field Laboratory. He
acknowledges the support by RFBR grant N 00-02-17729a and I.V. acknowledges
the support by Israel Science Foundation founded by the Israel Academy of
Sciences and Humanities.

\end{document}